\documentclass[10pt,prd,showpacs,twocolumn,psfig,epsfig,superscriptaddress,floatfix]{revtex4}
\usepackage[latin9]{inputenc}
\usepackage{amsmath}
\usepackage{amssymb}
\usepackage{graphicx}

\makeatletter
\@ifundefined{textcolor}{}
{%
 \definecolor{BLACK}{gray}{0}
 \definecolor{WHITE}{gray}{1}
 \definecolor{RED}{rgb}{1,0,0}
 \definecolor{GREEN}{rgb}{0,1,0}
 \definecolor{BLUE}{rgb}{0,0,1}
 \definecolor{CYAN}{cmyk}{1,0,0,0}
 \definecolor{MAGENTA}{cmyk}{0,1,0,0}
 \definecolor{YELLOW}{cmyk}{0,0,1,0}
}

\usepackage{dcolumn}
\usepackage{bm}
\usepackage{amsfonts}

\makeatother

\begin{document}

\title{Studying color connection effects of $e^+e^- \to c\bar{c}c\bar{c} \to \Xi_{cc}+X $ process within Quark Combination Model}

\author{Y. Jin}

\affiliation{School of Physics and Technology, University of Jinan, Jinan 250022,
People?s Republic of China}

\author{H. L. Li}

\affiliation{School of Physics and Technology, University of Jinan, Jinan 250022,
People?s Republic of China}

\author{S. Q. Li}

\affiliation{School of Physics, Shandong University, Jinan 250100, People?s Republic of China}

\author{S. Y. Li}

\affiliation{School of Physics, Shandong University, Jinan 250100, People?s Republic of China}

\author{Z. G. Si}

\affiliation{School of Physics, Shandong University, Jinan 250100, People?s Republic of China}
\affiliation{CCEPP, IHEP, Beijing 100049, People?s Republic of China}

\author{T. Yao}

\affiliation{School of Physics, Shandong University, Jinan 250100, People?s Republic of China}

\author{X. F. Zhang}

\affiliation{School of Physics and Technology, University of Jinan, Jinan 250022,
People?s Republic of China}
\affiliation{School of Physics, Shandong University, Jinan 250100, People?s Republic of China}

\begin{abstract}
The color connection of the parton system is an important bridge to connect the perturbative process and the hadronization one. 
The special color connection of four-heavy-quark system in $e^+e^-$ annihilation, which is the necessary one for the doubly heavy baryon and tetraquark productions is revisited. The hadronization effects, investigated with the help of the Quark Combination Model are compared with the corresponding results employing the Lund String Model in our previous work. 
The global properties related to a certain color connection are not sensitive to various  hadronization models.


\end{abstract}

\date{\today}\maketitle

In high energy collisions, hadronization is one of the most important processes for understanding the non-perturbative quantum chromodynamics and the confinement mechanism. $e^{+}e^{-}$ annihilation has the advantage that the hadronization results can be compared with the experimental data directly to extract informations of hadronization because of no hadrons in the initial states. To embed the parton system into a hadronization model, it is necessary to specify the concrete color connection. However, the color connection for  multi-parton system is not unique. 
Some clues can be obtained by analyzing the decomposition of the color space of the final partons \cite{WQEFH,Wagus,hanw,jin2,jin3}. One of the interesting and important examples is the four-heavy-quark system ($QQ'\bar{Q}\bar{Q}'$), in which many phenomena related to QCD properties can be studied, \textit{e.g.}, the (re)combination of quarks in the production of special hadrons, the influence of soft interaction on the reconstruction of intermediate particles (such as $W^{\pm}$), \textit{etc.}, most of which are more or less related to the color connections among these four quarks. 
 In Ref.~\cite{jin2}, the interesting decompositions of the color space of the $QQ'\bar{Q}\bar{Q}'$ system in $e^+e^-$ annihilation have been discussed, which are 
\begin{eqnarray}
(3_{Q}\otimes3_{Q'})\otimes3_{\bar{Q}}^{*}\otimes3_{\bar{Q}'}^{*}=3_{QQ'}^{*}\otimes3_{\bar{Q}}^{*}\otimes3_{\bar{Q}'}^{*}\oplus\cdots, & or\nonumber \\
3_{Q}\otimes3_{Q'}\otimes(3_{\bar{Q}}^{*}\otimes3_{\bar{Q}'}^{*})=3_{Q}\otimes3_{Q'}\otimes 3_{\bar{Q}\bar{Q}'}\oplus\cdots,
\label{q2qq}
\end{eqnarray}
where $3$ and $3^{*}$ denote the triplet and anti-triplet representations of the $SU_{c}(3)$ Group respectively, and the subscripts correspond to the relevant (anti)quark. We investigated the four-heavy-quark system in which only one $QQ'$ (or $\bar{Q}\bar{Q}'$) pair has a small invariant mass, while the invariant masses of the other two are left unrestricted, {\it i.e.}, a three-jet event configuration. When two quarks (antiquarks) with small invariant masses in the color state $3^{*}(3)$ attract each other and can be considered as a diquark (or an antidiquark), it can hadronize into a(n) (anti)baryon (tetraquark). The corresponding hadronization procedure is a `branching' process via the creation of quarks from vacuum by the strong interactions within the system. 
 Let's take $cc\bar{c}\bar{c}$ as an example, considering the concrete case that $(cc)$ combines with a quark $q$ (antidiquark $\bar{q}\bar{q}'$) to form $\Xi_{cc}$ \cite{majp,Jiang:2013ej} ($T_{cc}$ \cite{liuyanrui}).  
To balance the quantum numbers of color and flavor, an antiquark/diquark must be simultaneously created from vacuum. To branch them further, more
quark pairs and diquark pairs must be created from vacuum via the interactions among the quark system. Such a cascade process will proceed until the end of time when most of the `inner energy' of the entire system is transformed into the kinematical energies
and masses of the produced hadrons. 
Each of two newly created quarks (antidiquarks) combines with each of the primary $\bar{c}$ quarks to respectively hadronize into two open charmed hadrons, which can be described by an assigned concrete hadronization model (for details, see \cite{string,pythia,cluster,herwig}). 
We adopted the Lund String Model (LSM) \cite{string, pythia} in our previous work \cite{jin2}. In this paper, we employ the Quark Combination Model (QCM) \cite{Si:1997zs,Xie2} to deal with the hadronization and compare with the corresponding previous results in \cite{jin2} to study to what extent different hadronization models interplay with this kind of color connection. 
 As other hadronization models, QCM successes in reproducing experimental data of $e^+e^-\to h's$ and $pp(\bar{p}) \to h's$ processes. As a matter of fact, the baryon to meson ratio \cite{Adler:2003kg,Adams:2006wk} and constituent quark number scaling of elliptic flow $v_2$ \cite{v2data} measured at RHIC experiments can be naturally understood by QCM. 
 Hence, such comparisons of various hadronization models connected with the same color connection can shed light on different aspects of soft strong interactions reflected by various models.  
 
\begin{figure}[htb]
\centering
\scalebox{0.2}{\includegraphics{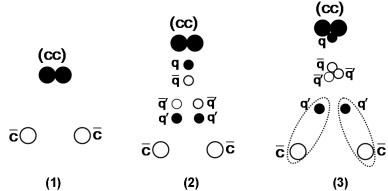}}
\caption{The procedure of $\Xi_{cc}$ production for the special color connection of Eq.~(\ref{q2qq}). Solid circles represent quarks, while hollow circles represent antiquarks. Each of two newly created quarks and one primary $\bar{c}$ are connected to form the color singlet system in (3).}
\label{4cfrag2}
\end{figure}
 
 The details of the application of QCM are similar to that of LSM in Ref.~\cite{jin2}. As shown in Fig.~\ref{4cfrag2}, {\it e.g.}, for $\Xi_{cc}$ production, the distribution of $\Xi_{cc}$ is described by the Peterson formula \cite{pet83} 
\begin{equation}
f(z)\propto\frac{1}{z(1-1/z-\epsilon_{Q}/(1-z))^{2}},\label{Peterson}
\end{equation}
where $\epsilon_{Q}$ is a free parameter which is expected to scale between
flavors as $\epsilon_{Q}\propto1/m_{Q}^{2}$. In the following, we will show the results corresponding to $\epsilon_{Q}=1/25$ for the $(cc)$ diquark. 
The complementary antiquark transits to an antibaryon by combining with an antidiquark and the distribution is described with the help of a standard fragmentation function \cite{Albino:2005me} as well as the analytical formulation employed in {\sc PYTHIA} \cite{pythia}
\begin{equation}
f(z)\propto z^{-1}(1-z)^{a}exp(-bm_{\perp}^{2}/z),\label{Anderson}
\end{equation}
where $a$ and $b$ are free parameters. In our program, for simplicity we use Eq.~(\ref{Anderson}) with $a=0.3$ GeV$^{-2}$ and $b=0.58$ GeV$^{-2}$ as in Ref.~\cite{pythia}. 
The balancing diquark is broken by the interactions within the remaining system, then each of two quarks becomes connected to the primary $\bar{c}$ quark to form two $q'\bar{c}$ separated  systems, and each is described by QCM \cite{Si:1997zs,Xie2} for the hadronization. 

In the framework of QCM, Quark Production Rule (QPR) and Quark Combination Rule (QCR) are adopted to describe the hadronization in a color singlet system, a simple Longitudinal Phase Space Approximation (LPSA) is used to obtain the momentum distribution for primary hadrons in their own system, and then this hadronization scheme is extended to the multi-parton states. 
For QPR, in a color singlet system formed by $q'\bar{c}$, the average number of the quark pairs $<N>$ produced by vacuum excitation is given by
\begin{equation}
\label{fir1}
<N>=\sqrt{\alpha^{2}+\beta(W-{{M}_{q'}}-{{M}_{\bar{c}}}+2\bar{m})} -\alpha-1,
\end{equation}
where $W$ is the invariant mass of the system, $\alpha=\beta\bar{m}-{{1} \over {4}}$, $\beta$ is a free parameter, $\bar{m}$ is the average mass of newborn quarks, and $M_{q'}$ and $M_{\bar{c}}$ are the masses of endpoint quark and antiquark.

For the combination process, all kinds of hadronization models satisfy the near correlation in
rapidity more or less. In Ref.~\cite{Xie1}, we have shown
that the nearest correlation in rapidity is in agreement with the
fundamental requirements of QCD, and determines QCR completely.
The rule guarantees that the combination of quarks across more
than two rapidity gaps never emerges and that $N$ quarks and $N$
antiquarks are exactly exhausted, and thus the unitarity is guaranteed  (see Ref.  \cite{Han:2009jw}). QCM treats meson and baryon formation uniformly, so the production ratio of baryon to meson can be directly obtained as
\begin{equation}
\label{w5}
{R_{B/M} =} {{(1-a)N -b} \over {3(aN+b)}},
\end{equation}
where $a=0.66$ and $b=0.56$. 

As in other phenomenological models, some inputs must be given to determine the momentum distribution of primary hadrons.
In this paper, we simply adopt the widely used LPSA which is equivalent to the
constant distribution of rapidity. Hence a primary hadron $i$ is
uniformly distributed on the rapidity axis, and its rapidity can be
written as
\begin{equation}
Y_{i}=Z+\xi_{i} Y, ~~~~
0 \leq \xi_{i} \leq 1,
\end{equation}
where $\xi_i$ is a random number;
$Z$ and $Y$ are two arguments, and can be determined by
energy-momentum conservation in such a color singlet system
\begin{equation}
{\sum \limits_{i=1}^{H}} E_{i} = W, ~~~~
{\sum \limits_{i=1}^{H}} P_{Li} = 0,
\end{equation}
where $E_{i}$ and $P_{Li}$ denote the energy and the longitudinal
momentum of the $i$th primary hadron respectively, obtained by
\begin {equation}
\left \{
\begin{array}{ll}
E_{i} = m _{Ti} {{exp(Y_{i}) + exp(-Y_{i})} \over {2}}\\
P_{Li}= m _{Ti} {{exp(Y_{i}) - exp(-Y_{i})} \over {2}}.
\end{array}
\right.
\end{equation}
where $m_{Ti}$ is given by
\begin{equation}
m _{Ti} = \sqrt{m_{i}^2 + {\stackrel {\rightarrow} P_{Ti}}^{2}},
\end{equation}
where $m_{i}$ is the mass of the $i$th primary hadron,
and $\stackrel {\rightarrow} P_{Ti}$ obeys the distribution
\begin{equation}
\label{last}
f({\stackrel{\rightarrow} P_{T1}},\ldots,
{\stackrel{\rightarrow}P_{TH}}) \propto {\prod \limits_{i=1}^{H}}
exp(-{{{\stackrel {\rightarrow} P_{Ti}}^{2}} \over {\sigma^{2}}})
\delta ({\sum \limits_{i=1}^{H}} {\stackrel {\rightarrow} P_{Ti}}).
\end{equation}
In this paper, we set $\sigma=0.2~$GeV. Eq.~(\ref{last}) is
just what LSM uses.

\begin{figure}[htb]
\centering
\begin{tabular}{cccc}
\scalebox{0.15}{\includegraphics{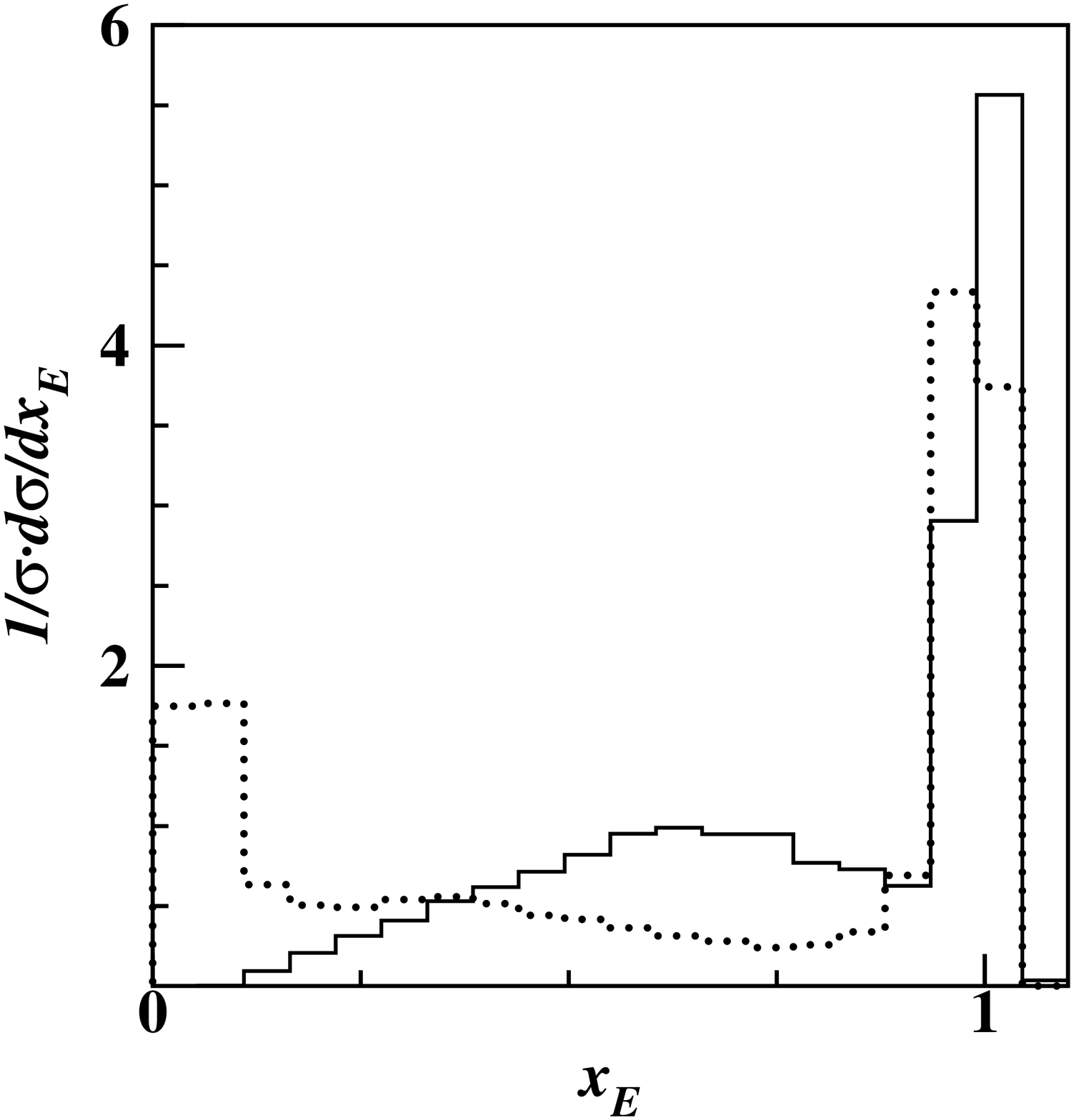}}&
\scalebox{0.15}{\includegraphics{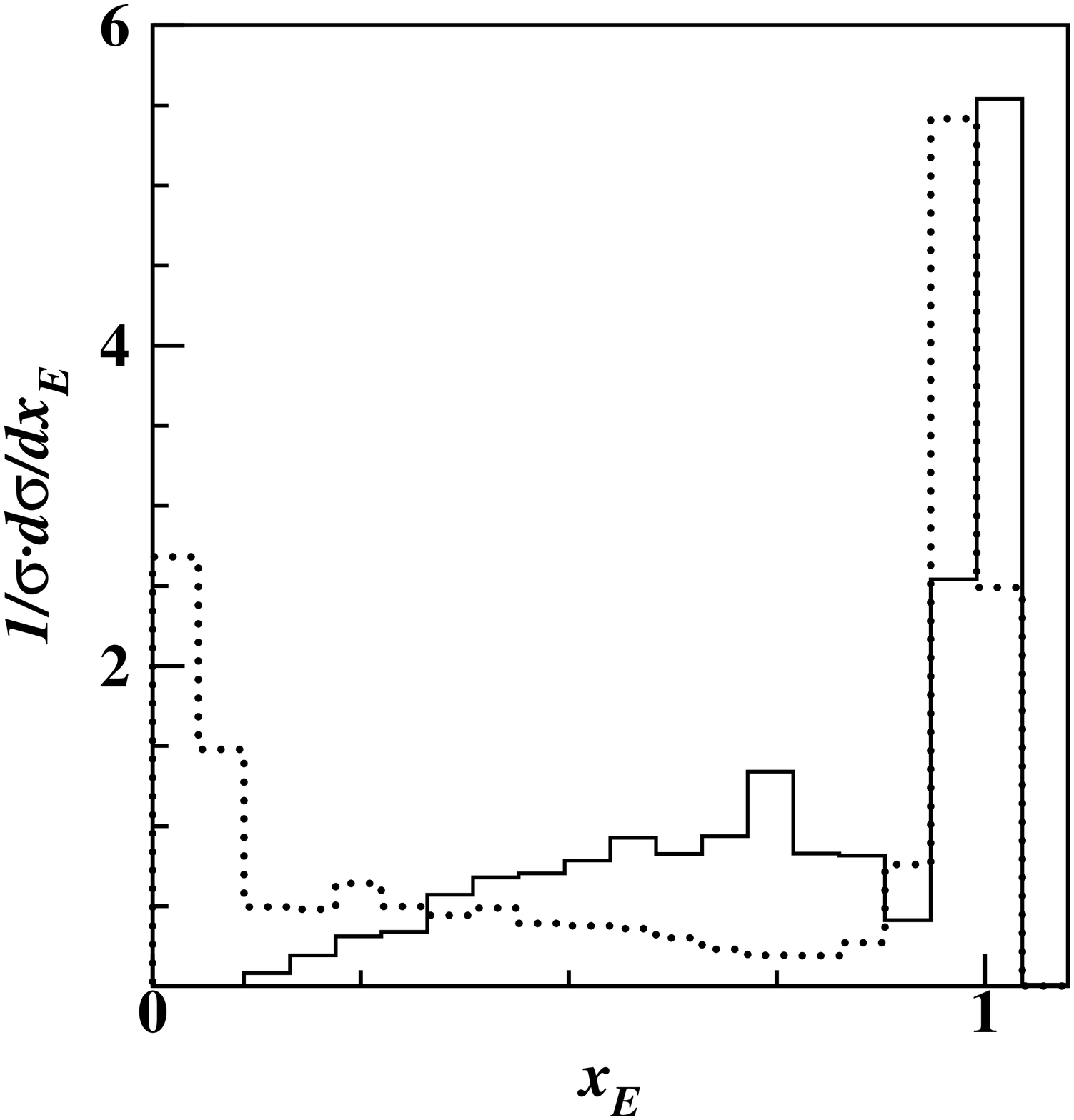}}&\\
{\scriptsize (a)}&{\scriptsize (b)}
\end{tabular}
\caption{
The distribution of the energy fraction for the jet involving $\Xi_{cc}$ compared to the jet  from the primary $\bar{c}$ as a function of the scaled dimensionless variable $x_{E}=2E/\sqrt{s}$. The solid line represents the jet involving $\Xi_{cc}$, and the dotted line represents the jet from the primary $\bar{c}$. (a)/(b) stands for QCM/LSM.}
\label{xejet}
\end{figure}

\begin{figure}[htb]
\centering
\begin{tabular}{cccc}
\scalebox{0.14}{\includegraphics{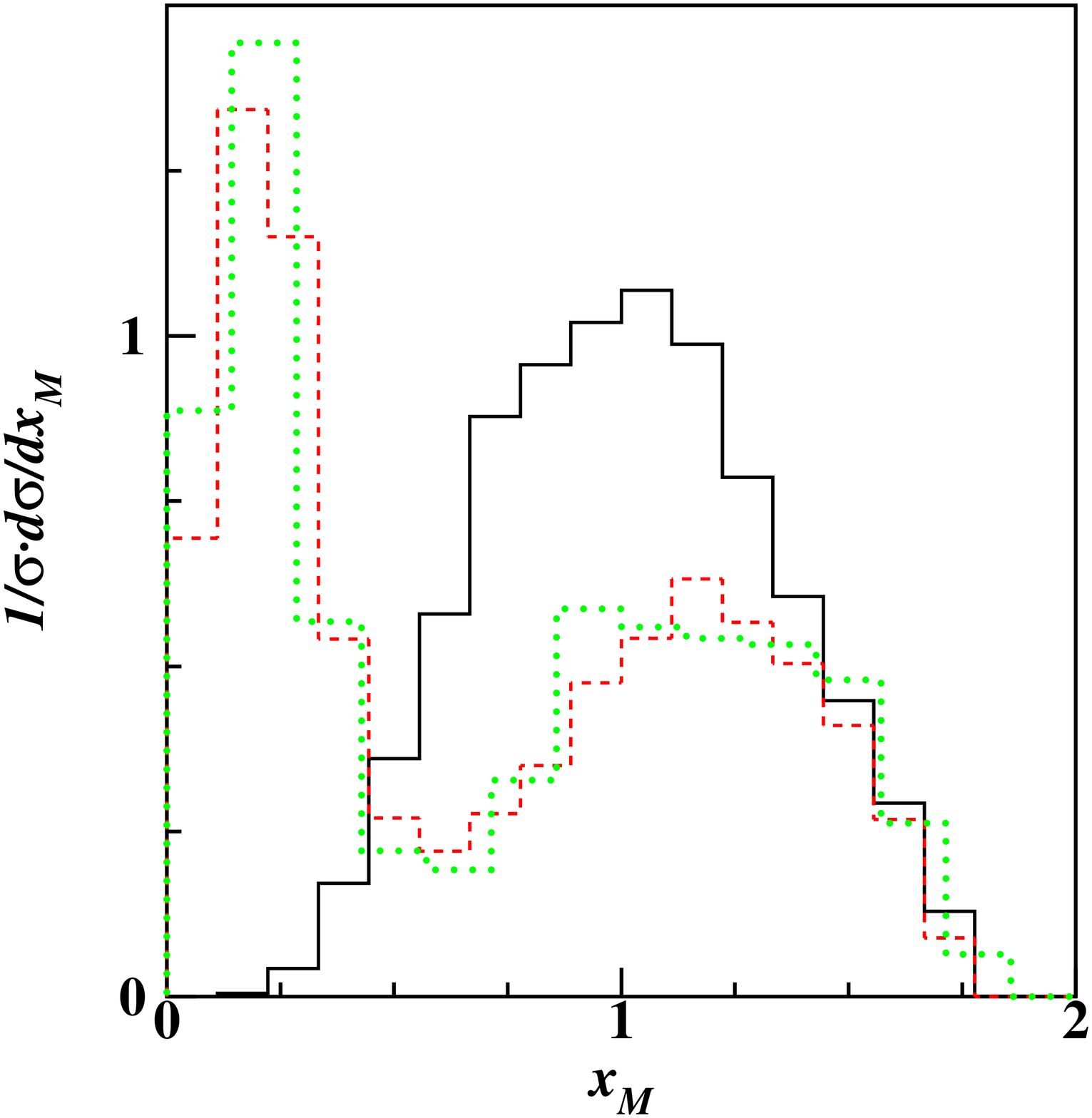}}&
\scalebox{0.16}{\includegraphics{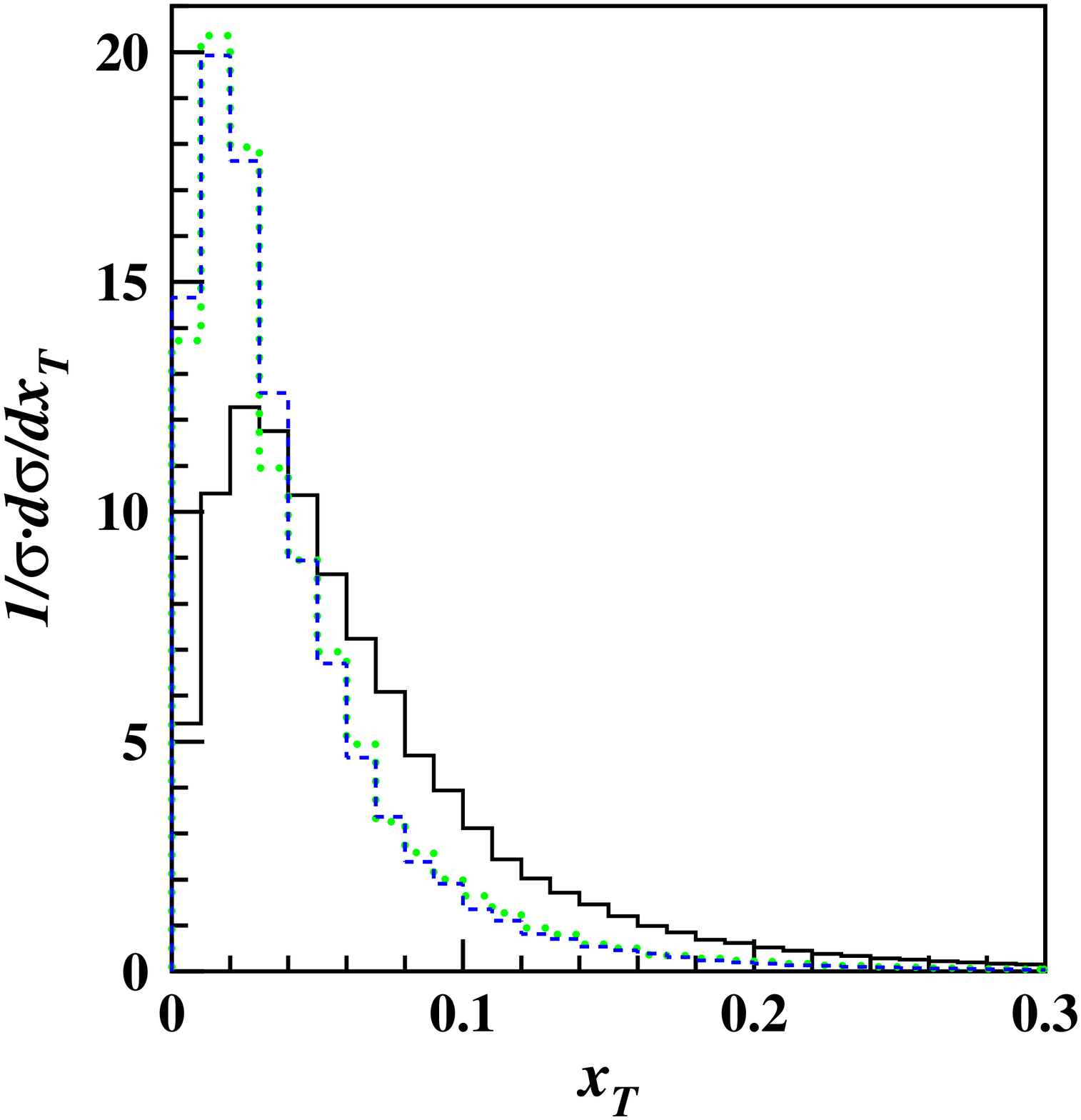}}&\\
{\scriptsize (a)}&{\scriptsize (b)}
\end{tabular}
\caption{
(a) The distribution for the invariant mass of two primary $\bar{c}$ quarks compared to two primary $\bar c$ jets system as a function of the scaled dimensionless variable $x_{M}=2M_{inv}/\sqrt{s}$. The solid black line represents two $\bar{c}$ quarks, and the dashed red (dotted green) line represents two primary $\bar c$ jets system for QCM (LSM); 
(b) The transverse-momentum distribution for $\Xi_{cc}$ compared to the diquark $(cc)$ as a function of the scaled dimensionless variable $x_{T}=2p_T/\sqrt{s}$. The solid black line represents $(cc)$, and the dashed blue (dotted green) line represents $\Xi_{cc}$ for QCM (LSM).}
\label{xpt}
\end{figure}

For the numerical results, similarly to \cite{jin2}, the center-of-mass energy is set to $Z^0$ pole. 
At first, we investigate the hadronization effects on jets. Here we take JADE algorithm \cite{JADE} as $y_{ij}=\frac{(p_i+p_j)^2}{E_{cm}^{2}}$ to define jets. The parameter $y_{cut}$ is thus introduced and two partons/particles are considered as being in one jet when $y_{ij}<y_{cut}$. We apply the jet algorithm to the parton level to obtain the three-jet events of $e^+e^- \to (cc)\bar{c}\bar{c}$. Hereafter, three-jet events are defined with $y_{cut}=10^{-3}$ if no explicit explanation is given.
 Fig.~\ref{xejet} shows the energy fraction distribution of the jet involving $\Xi_{cc}$ and that of the jet from the primary $\bar{c}$. It is obvious that the jet involving $\Xi_{cc}$ is the hardest one. Both QCM and LSM give the similar results. 
In the hadronization process, the energy and momentum are broadened relative to those of the parton level. To show this effect, the distribution for the invariant mass of two primary $\bar c$ jets system is displayed in Fig.~\ref{xpt}(a). The distribution spreads more widely than that of two primary $\bar{c}$ quarks.

To further demonstrate the hadronization effects, the transverse-momentum with respect to the thrust axis distributions of $\Xi_{cc}$ are shown in Fig.~\ref{xpt}(b). The thrust axis is determined from the final hadrons system. The absolute values of the distributions more or less depend on the parameters in Eq.~(\ref{Peterson}), (\ref{Anderson}) and the entire hadronization. Once data are available at the future relevant experiments, {\it e.g.}, International Linear Collider, the Higgs factory and $Z^0$ factory, {\it etc.}, the fragmentation functions and the parameters can be tuned according to the comparisons with data. From Fig.~\ref{xpt}(b), one can also find that there is little difference between the results from QCM and LSM. 

\begin{figure}[htb]
\centering
\begin{tabular}{cccc}
\scalebox{0.15}{\includegraphics{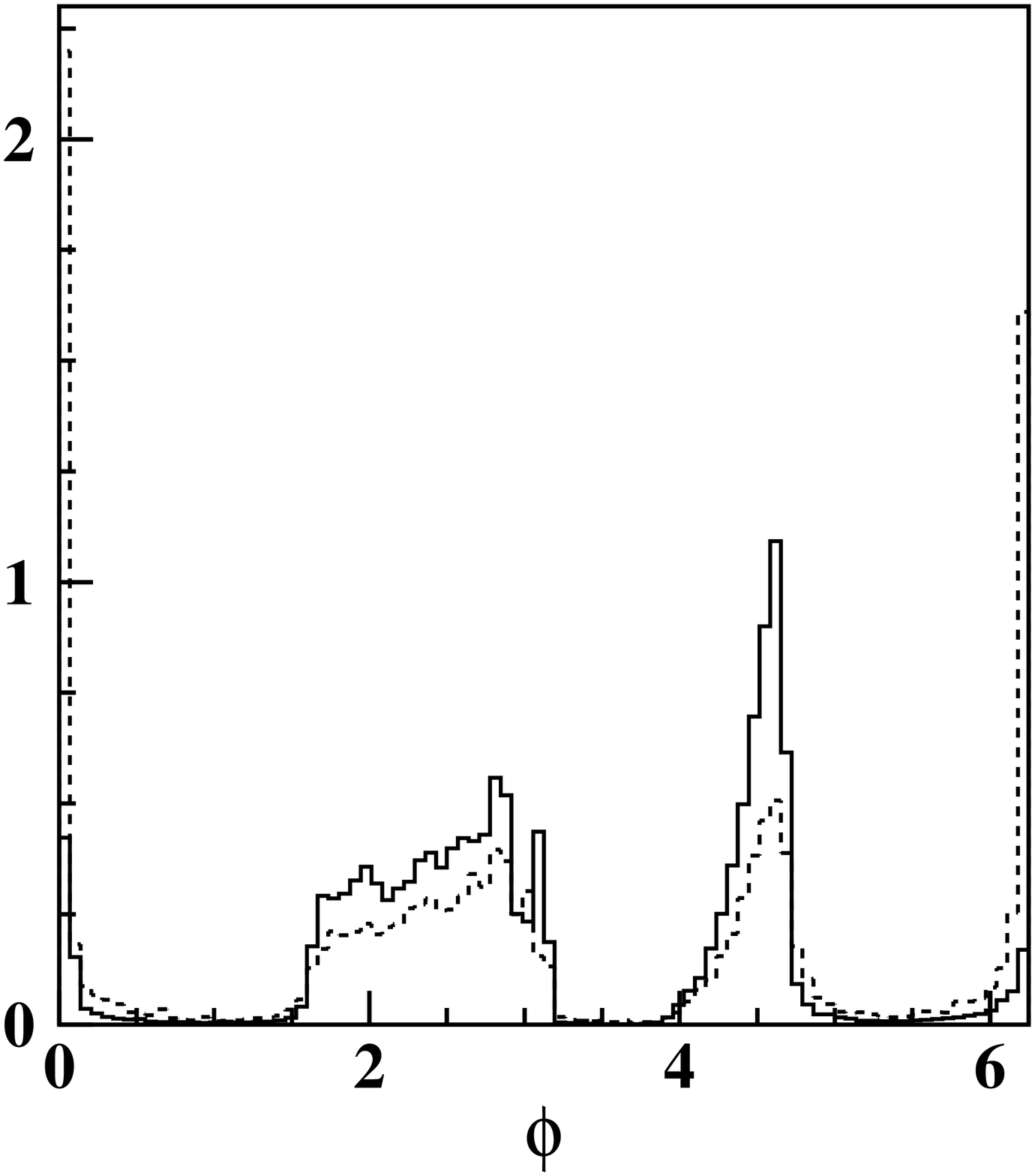}}&
\scalebox{0.15}{\includegraphics{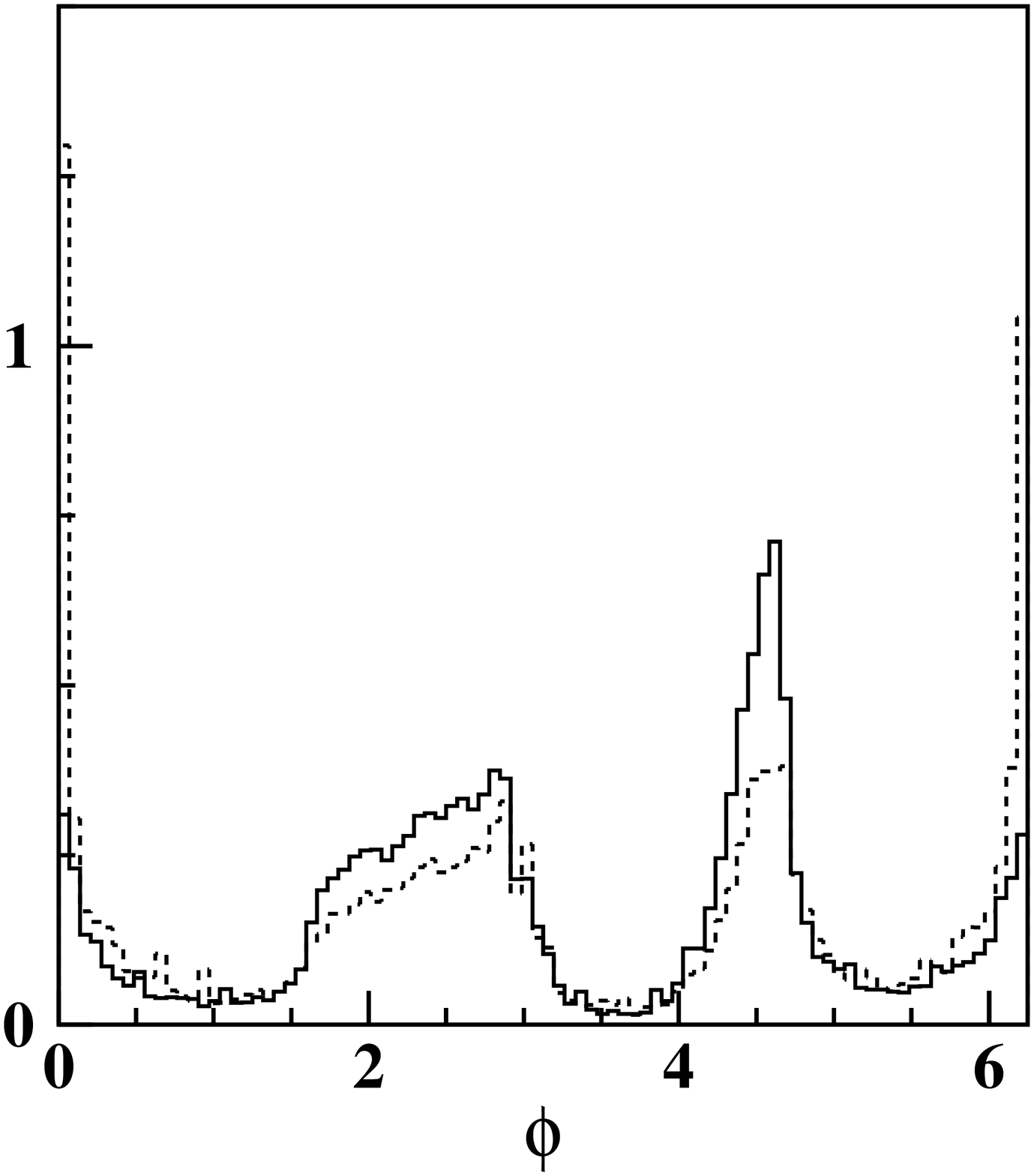}}&\\
{\scriptsize (a)}&{\scriptsize (b)}
\end{tabular}
\caption{
(a) The solid (dashed) line represents $\frac{1}{N}\frac{dN}{d\phi}$ for QCM (LSM); 
(b) The solid (dashed) line represents $\frac{1}{E}\frac{dE}{d\phi}$ for QCM (LSM). 
The symmetric three-jet events are defined here to require each relative angle to be larger than $\frac{\pi}{2}$.}
\label{hnphi}
\end{figure}

In the special color connection discussed in this paper, it can be clearly seen from Fig.~\ref{4cfrag2} that there is no color flow between two primary $\bar{c}$ quarks. So after hadronization, there should be few hadrons emerging in the phase-space between them. This is the so-called `string effect' and the corresponding event shape should not be sensitive to which hadronization model is employed. This is clearly demonstrated by what is shown in Fig.~\ref{hnphi}. 
Here we choose the more symmetric three-jet events by requiring that the angle between any two jets is larger than $\pi/2$. 
Because of momentum conservation, all three jet momenta must be in the same plane (${\cal P}$) in the $e^{+}e^{-}$ center-of-mass frame \cite{Ellis:1976uc}. The three-momentum of each final-state particle $\vec{k}_i$ is projected onto one of the three regions between the jets to obtain the two-dimensional vector $\vec{k}_i'$ in the plane ${\cal P}$. The three-momentum of the jet that contains $\Xi_{cc}$ is chosen to be the $x$ axis. The angle between $\vec{k}_i'$ and the $x$ axis is the azimuthal angle $\phi$ of the corresponding particle. 
We can then calculate the final particle-number (energy) distribution $1/N\,dN/d\phi$ ($1/E\,dE/d\phi$). The corresponding results are shown in Fig.~\ref{hnphi}. Obviously, both QCM and LSM can give this kind of effect.

The above distributions of the transverse-momentum and final particle-number (energy) as well as the string effect which is the fingerprint of the fact that there is no color flow between the separate primary heavy antiquarks, are found determined by the color configuration and not sensitive to the hadronization models. 
Therefore the study of color connections is important for understanding the non-perturbative quantum chromodynamics. The future relevant experiments, {\it e.g.}, International Linear Collider, the Higgs factory and $Z^0$ factory, {\it etc.}, can provide opportunities for this study. 
Moreover, it is well known that the hadronization models should be universal for different hadronization processes. So QCM and LSM, {\it etc.}, can also be applied in other hadronization processes, {\it e.g.}, $pp$ collision in Large Hadron Collider (LHC) and DIS. 
 For the case that four-quark system emerges which has large rapidity gap with other clusters, the hadronization effects can also be studied by QCM when suitable observables introduced, {\it e.g.}, the ratio of baryon to meson of Eq.~(\ref{w5}). In addition, $(cc)$ can be regarded as a colored cluster, so if similar heavy colored particles \cite{Kats:2012ym,Chen:2014haa}  beyond Standard Model, {\it e.g.}, hidden valley quarks \cite{Han:2007ae, Essig:2009nc}, are produced at LHC, our method can provide useful hints to investigate the related hadronization phenomena.

\section*{Acknowledgments}

We dedicate this paper to the memory of Prof. Qu-Bing Xie  and Prof. Xi-Ming Liu for their initialization works of 
the Quark Combination Model employed here. This work is supported in part by the Natural Science Foundation of  Shandong Province (ZR2014AM016, ZR2013AQ006, JQ201101) and National Natural Science Foundation of China (11325525, 11275114, 11305075).


\begin{thebibliography}{150}


\bibitem{WQEFH} Qun Wang and Qu-Bing Xie, 
 Phys.\ Rev.\ D \textbf{52}, 1469, 1995. 



\bibitem{Wagus} Qun Wang, G. Gustafson and Qu-Bing Xie, 
 Phys.\ Rev.\ D \textbf{62} 054004, 2000. 



\bibitem{hanw} W.~Han, S.~-Y.~Li, Z.~-G.~Si and Z.~-J.~Yang,
 Phys.\ Lett.\ B \textbf{642}, 62 (2006). 



\bibitem{jin2} 
Y.~Jin, S.~-Y.~Li, Z.~-G.~Si, Z.~-J.~Yang and T.~Yao,
  Phys.\ Lett.\ B {\bf 727}, 468 (2013).


\bibitem{jin3} 
  Y.~Jin, S.~Y.~Li, Y.~R.~Liu, Z.~G.~Si and T.~Yao,
  Phys.\ Rev.\ D {\bf 89}, no. 9, 094006 (2014).
  
  
  \bibitem{majp} 
    J.~P.~Ma and Z.~G.~Si,
    Phys.\ Lett.\ B {\bf 568} (2003) 135.
    
    
  \bibitem{Jiang:2013ej}
    J.~Jiang, X.~-G.~Wu, S.~-M.~Wang, J.~-W.~Zhang and Z.~-Y.~Fang,
    Phys.\  Rev.\  D 87, {\bf 054027} (2013).
  
  
  
  \bibitem{liuyanrui} S. Zouzou, B. Silvestre-Brac, C. Gignoux, J. M. Richard, Z. Phys. C \textbf{30}, 457 (1986).
  
  
  

\bibitem{string} B. Andersson, `The Lund Model' (Cambridge University Press, 1998), and references therein.


\bibitem{pythia}
T.~Sjostrand, P.~Eden, C.~Friberg, L.~Lonnblad, G.~Miu, S.~Mrenna and E.~Norrbin,
  Comput.\ Phys.\ Commun.\  {\bf 135}, 238 (2001).
   
   
   \bibitem{cluster} G.~Marchesini and B. R. Webber, 
    Nucl.\ Phys.\ B \textbf{238}, 1, 1984; 
    B.~R.~Webber, 
    Nucl.\ Phys.\ B \textbf{238}, 492, 1984. 
   
   
   
   \bibitem{herwig} G. Corcella et al., JHEP 0101, 010, 2001.



  \bibitem{Si:1997zs}
  Si Z G, Xie Q B, Wang Q.
   Commun Theor Phys, 1997, 28: 85-94;
  Si Z G, Xie Q B.
   High Ener Phys and Nucl Phys, 1999, 23: 445-458.
   
   
   
   
  \bibitem{Xie2}
  Xie Q B, Liu X M.
   Quark production rule in $e^+e^-$ annihilation to two jets.
   Phys Rev D, 1988, 38: 2169-2177



\bibitem{Adler:2003kg}
Adler SS {\it et al.},
Phys.\ Rev.\ Lett. {\bf 91}, 172301 (2003).



\bibitem{Adams:2006wk}
  Adams J {\it et al.},
  arXiv:nucl-ex/0601042;
  Long H,
  J.\ Phys.\ G {\bf 30}, 193 (2004).
  


\bibitem{v2data}
S.~S.~Adler {\it et al.}  [PHENIX Collaboration],  Phys.\ Rev.\
Lett.\  {\bf 91}, 182301 (2003);
J.~Adams {\it et al.}  [STAR Collaboration], Phys.\ Rev.\ Lett.\
{\bf 92}, 052302 (2004);
J.~Adams {\it et al.}  [STAR Collaboration], Phys.\ Rev.\ Lett.\
{\bf 95}, 122301 (2005).





\bibitem{pet83} C. Peterson, D. Schlatter, I. Schmitt and P. Zerwas,
Phys. Rev. D \textbf{27} (1983) 105.



\bibitem{Albino:2005me} 
  S.~Albino, B.~A.~Kniehl and G.~Kramer,
  Nucl.\ Phys.\ B {\bf 725}, 181 (2005).




\bibitem{Xie1}
Xie Q B, in Proceedings of the XIXth International Symposium on
Multi-particle Dynamics 1988, edited by D. Schiff and J. Tran
Thanh Van (World Scientific, Singapore 1988), P. 369,
and the references therein.

 
 
\bibitem{Han:2009jw}
  W.~Han, S.~-Y.~Li, Y.~-H.~Shang, F.~-L.~Shao and T.~Yao,
  Phys.\ Rev.\ C {\bf 80} (2009) 035202.
  
  
  \bibitem{JADE}JADE Collaboration, W. Bartel et al., Z. Phys. C {\bf33}, 23 (1986); S. Bethke, Habilitation thesis, LBL 50-208 (1987).
  
  
  
   \bibitem{Ellis:1976uc} J.~R.~Ellis, M.~K.~Gaillard and G.~G.~Ross,
    Nucl.\ Phys.\ B \textbf{111} (1976) 253 {[}Erratum-ibid.\ B \textbf{130}
   (1977) 516{]}.
   
   
  
\bibitem{Kats:2012ym} 
  Y.~Kats and M.~J.~Strassler,
  JHEP {\bf 1211}, 097 (2012).
  
  
  \bibitem{Chen:2014haa} 
    C.~Y.~Chen, A.~Freitas, T.~Han and K.~S.~M.~Lee,
    arXiv:1410.8113 [hep-ph].
  
  
 
 
  \cite{Han:2007ae}
  \bibitem{Han:2007ae} 
    T.~Han, Z.~Si, K.~M.~Zurek and M.~J.~Strassler,
    JHEP {\bf 0807}, 008 (2008).
    
    
    \bibitem{Essig:2009nc} 
      R.~Essig, P.~Schuster and N.~Toro,
      Phys.\ Rev.\ D {\bf 80}, 015003 (2009).
   
   
 

\end{thebibliography}
\end{document}